\documentclass[fleqn,10pt]{wlscirep}
\usepackage[utf8]{inputenc}

\usepackage{dcolumn}
\usepackage{bm}

\usepackage[english]{babel}
\usepackage{amssymb,amsfonts,amsmath,mathtools,mathrsfs}
\usepackage{relsize}
\usepackage{mdframed}
\usepackage{enumitem}
\usepackage{graphicx}
\usepackage{float}
\usepackage{url,hyperref}
\usepackage{gensymb}
\usepackage{grffile}

\usepackage[bottom]{footmisc}
\usepackage[capitalize]{cleveref}
\usepackage{textcomp}
\usepackage{braket} 

\newcommand{\upa}{\ensuremath{\uparrow}}
\newcommand{\dna}{\ensuremath{\downarrow}}
\newcommand{\la}{\ensuremath{\langle}}
\newcommand{\ra}{\ensuremath{\rangle}}
\newcommand{\lv}{\ensuremath{\lvert}}
\newcommand{\rv}{\ensuremath{\rvert}}

\newcommand{\eps}{\ensuremath{\epsilon}}
\newcommand{\summ}[1]{\sum\limits_{#1}}

\newcommand{\mbf}[1]{\ensuremath{\mathbf{#1}}}

\newcommand{\vk}{\ensuremath{\mathbf{k}}}

\newcommand{\vecr}{\ensuremath{\mathbf{r}}}

\newcommand{\ham}{\ensuremath{\mathcal{H}}}
\newcommand{\bdelta}{\ensuremath{\boldsymbol{\delta}}}
\newcommand{\mcal}[1]{\ensuremath{\mathcal{#1}}}

\title{Pairing symmetries in the Zeeman-coupled extended attractive Hubbard model}

\author[1,2]{Swagatam Nayak}
\author[1,3]{Navketan Batra}
\author[1,*]{Sanjeev Kumar}
\affil[1]{Department of Physical Sciences, Indian Institute of Science Education and Research Mohali, Mohali, Manauli PO 140306, India}
\affil[2]{School of Physical Sciences, National Institute of Science Education and Research (NISER), Bhubaneswar, Odisha 752050, India}
\affil[3]{Department of Physics, Brown University, Providence, RI 02912, USA}

\affil[*]{sanjeev@iisermohali.ac.in}

\begin{abstract}

By introducing the possibility of equal- and opposite-spin pairings concurrently, we show that the extended attractive Hubbard model (EAHM) exhibits rich ground state phase diagrams with a variety of singlet, triplet, and mixed parity superconducting orders. We study the competition between these superconducting pairing symmetries invoking an unrestricted Hartree-Fock-Bogoliubov-de Gennes (HFBdG) mean-field approach, and we use the $d$-vector formalism to characterize the nature of the stabilized superconducting orders. We discover that, while all other types of orders are suppressed, a non-unitary triplet order dominates the phase space in the presence of an in-plane external magnetic field. We also find a transition between a non-unitary to unitary superconducting phase driven by the change in average electron density. Our results serve as a reference for identifying and understanding the nature of superconductivity based on the symmetries of the pairing correlations. The results further highlight that EAHM is a suitable effective model for describing most of the pairing symmetries discovered in different materials.

\end{abstract}
\begin{document}

\flushbottom
\maketitle

\thispagestyle{empty}

\section{Introduction}
\label{sec_Introduction}

In a conventional $s$-wave superconductor, described by BCS-Migdal-Eliashberg theory\cite{cooper1956bound,bardeen1957theory,bardeen1957microscopic,migdal1958interaction,eliashberg1960interactions,eliashberg1961temperature}, a phonon-mediated effective attraction causes the electrons to form spin-singlet (with total spin $S=0$) Cooper pairs with  an isotropic $s$-wave orbital order parameter (OP) symmetry\cite{tinkham2004introduction}. Most of the superconductors discovered in the early phase of the past century have this type of conventional OP symmetry\cite{bennemann2008history}. On the contrary, the general consensus about the novel high $T_{c}$ cuprate superconductors is that they are identified as strong candidates for unconventional $d$-wave superconductors\cite{coffey1993quasiparticle,tsuei2000phase,oates2004observation}, which support the formation of spin-singlet Cooper pairs with an anisotropic $d$-wave orbital OP symmetry. In general, Cooper pairs can also be formed in the spin-triplet state, with total spin $S=1$ and anisotropic orbital OP\cite{maki2009triplet,mackenzie2003superconductivity}. Recent studies revealed that a two dimensional chiral $p$-wave spin-triplet superconductor is a candidate to host Majorana fermions \cite{kopnin1991mutual,volovik1999fermion,sarma2006proposal}. A pair of Majorana fermions, bound to topological defects, together known as Ising anyons, exhibit non-Abelian exchange statistics and such an object can be considered to be the potential building block for decoherence free quantum computation\cite{beenakker2013search}. Regardless of their technological relevance, superconductors with triplet-superconductivity emerging due to the intrinsic properties of the material itself are quite rare in nature. While $^{3}\text{He}$ was the first charge-less many body system where triplet-pairing state was realised\cite{thouless1960perturbation,brueckner1960level,anderson1961generalized,leggett1975theoretical}, it was $\text{UPt}_{3}$ which was identified as the first spin-triplet superconductor in charged many body systems\cite{tou1996odd,tou1998nonunitary}. Strong candidate materials for triplet-superconductivity, so far, include Uranium based heavy-fermion superconductors $\text{UPt}_{3}$, $\text{UGe}_{2}$, URhGe, UCoGe\cite{aoki2019review}, organic superconductor $\text{(TMTSF)}_{2}\text{PF}_{6}$\cite{lee2001triplet}, the promising perovskite superconductor $\text{Sr}_{2}\text{RuO}_{4}$\cite{ishida1998spin,maeno1999experimental}, and probably recently discovered heavy-fermion superconductor $\text{UTe}_{2}$\cite{ran2019nearly,metz2019point,sundar2019coexistence}. 

In conventional superconductors time reversal symmetry (TRS) allows energetically degenerate electron states $\lv\vk\upa\ra$ and $\lv-\vk\dna\ra$ to form spin-singlet Cooper pairs. On the other hand, parity symmetry (inversion symmetry in this scenario) along with TRS is needed to form odd-parity spin-triplet Cooper pairs\cite{sigrist1987symmetry,tsuei2000pairing,annett1995unconventional,annett1990symmetry}.
In case of broken inversion symmetry, there is also the possibility of forming Cooper pairs with mixed-parity superconducting (SC) states\cite{sigrist1987symmetry,gor2001superconducting,annett1990symmetry}. Possible occurrence of mixed parity SC states have been reported in several theoretical studies on different contexts, including quantized vortices of superfluid $^{3}\text{He}$\cite{salomaa1987quantized}, helical mixed parity SC phase relevant to the unconventional SC phases of $\text{UPt}_{3}$\cite{mineev1994helical}, Larkin-Ovchinnikov-Fulde-Ferrell phase with inhomogeneous order parameter\cite{romano2010field,matsuo1994order}, two-dimensional superconductors with broken inversion symmetry\cite{gor2001superconducting}, noncentrosymmetric superconductors\cite{fujimoto2009unambiguous}, vortex phase of a $d$-wave superconductor in presence of paramagnetic effects\cite{lebed2006cooper}, SC topological insulators in presence of surface Dirac fermions\cite{mizushima2014dirac}, and disordered monolayer transition metal dichalcogenides\cite{mockli2018robust}. While the search for unconventional pairing mechanism has not been an easy path, understanding unconventional pairing symmetries of the OP has been of utmost importance since it certainly gives us clues about the possible pairing mechanisms. Therefore, understanding and characterizing the unconventional OP symmetries of different spin-triplet and mixed-parity SC states remains a problem of crucial importance.

Most interestingly, the response of a superconducting state to an external magnetic field can reveal significant details about the pairing state of the Cooper pairs. For example, unlike the fully gapped $s$-wave superconductors, $d_{x^2-y^2}$-wave superconductors exhibit gapless quasiparticle spectrum, which allows even a weak amount of Zeeman field to spin-polarize the quasiparticle excitations, resulting in a stronger response to an external magnetic field\cite{ekin1997correlation,kashiwaya2006zeeman,prohammer1990upper}. For a two dimensional system it is quite interesting to study the effect of an in-plane (parallel to the plane) magnetic field, as the orbital motions of the electrons do not couple with the external magnetic field at all, and the Zeeman coupling of the electronic spins to the magnetic field is enough to study the response of the superconducting state. The effect of an in-plane magnetic field has been studied in various contexts, including quasi-2D systems like cuprates, emergence of a dissipative state in a superconducting $\text{Mo}_{0.79}\text{Ge}_{0.21}$ nanostrip, anisotropy of the upper critical field in $\text{Sr}_{2}\text{RuO}_{4}$, and magnetic field driven nodal topological superconductivity in monolayer transition metal dichalcogenides\cite{wang2017parallel,he2018magnetic}.

In this article we provide a detailed study of the competition between different superconducting pairings and their response to non-zero temperature and Zeeman field. We consider an extended attractive Hubbard model, with on-site and nearest neighbour attractive interaction on a square lattice, and treat the many-body interaction terms using mean-field analysis. We decouple the interaction terms into pairing channels without imposing any symmetry constraint on the superconducting pairing correlations. By doing so, it allows one to compare different superconducting ground states with different symmetries and spin pairings, i.e. a singlet and three components of triplet states. We employ the $d$-vector definition of the triplet order parameter to keep track of the triplet pairings throughout our analysis. We demonstrate that in this model numerous superconducting states with various symmetries can be stabilized and studied. Therefore, the extended attractive Hubbard model can be a universal effective model for studying unconventional, particularly triplet, superconductivity. The rest of the article is organized as follows. We start by presenting the model and defining all the parameters involved, and subsequently discussing the results: In section \ref{sec:RelativePhase}, using energy minimization arguments we self-consistently reduce the exploration parameter space and fix certain relative phase angles between different pairing correlations. Followed by section \ref{sec:competetionTripletStates}, in which we explore the competition between the three kinds of triplet pairings at non-zero magnetic field to lift the spin degeneracy. In section \ref{sec:GroundStatePD}, we present the ground state phase diagram of nearest-neighbour attractive strength versus particle density. Next in section \ref{sec:FiniteTempStudy} and \ref{sec:ZeemanCoupling}, we study the behaviour of different superconducting states at non-zero temperatures and Zeeman field respectively. Here we will encounter how different triplet and singlet orders react to the Zeeman field. Then we do a further characterization of superconducting phases that are stabilized in our study using symmetry and $d$-vector approach in section \ref{sec:Characterization}. Finally, we conclude by summarising our findings. All the methods we employ in this work, to analyse the model, are presented in the methods section at the end of the article. In the same section, we also discuss the $d$-vector formalism and its usefulness in characterizing triplet order parameters.

\section{Results and Discussions}

For superconductors like high-temperature Cuprates, the electronic band is quite narrow i.e., the orbitals have a small overlap between adjacent atoms. We therefore make use of the tight-binding Hamiltonian, which is either constructed from atomic orbitals or from Wannier orbitals, to study narrow band behaviours arising from electronic correlation effects. We, therefore, consider the Extended Attractive Hubbard Model, defined on a square lattice to be the doorway to address our problem. In real space, the Hamiltonian corresponding to the Extended Attractive Hubbard Model on a two-dimensional square lattice can be written as,

\begin{equation}
    \ham = \ham_{\text{kin}} + \ham_{\mu} + \ham_{\text{int}}^{\text{onsite}} + \ham_{\text{int}}^{\text{nn}} + \ham_{\text{B}},
\label{hamiltonian_total_compact}
\end{equation}

\noindent
where,
$\ham_{\text{kin}}$ represents the kinetic energy of the electrons, $\ham_{\mu}$ is the chemical potential term, $\ham_{\text{int}}^{\text{onsite}}$ is the on-site attractive interaction between two different spin projections, $\ham_{\text{int}}^{\text{nn}}$ is the nearest-neighbor (nn) attractive interaction which would be responsible for all the unconventional superconducting phases, and finally $\ham_{\text{B}}$ contains the Zeeman coupling of electron spin to an external magnetic field $B$. The corresponding operator terms, written in the second quantization notation are given as,

\begin{align}
    \ham_{\text{kin}} &= -t\summ{\la ij\ra,\sigma}c_{i\sigma}^{\dagger}c^{}_{j\sigma} + \text{H.c.}; ~~~~~~
    \ham_{\mu} = -\mu\summ{i\sigma}n_{i\sigma}
            = -\mu\summ{i\sigma}c_{i\sigma}^{\dagger}c^{}_{i\sigma}; ~~~~~~
    \ham_{\text{int}}^{\text{onsite}} = -U\summ{i}n_{i\upa}n_{i\dna} 
        = -U\summ{i}c_{i\upa}^{\dagger}c^{}_{i\upa} c_{i\dna}^{\dagger} c^{}_{i\dna} \notag\\
    \ham_{\text{int}}^{\text{nn}} &= -V\summ{\la ij\ra}n_{i}n_{j} 
        = -V\summ{\stackrel{\la ij\ra}{\sigma,\sigma^{\prime}}} 
        c_{i\sigma}^{\dagger}c^{}_{i\sigma} 
        c_{j\sigma^{\prime}}^{\dagger}c^{}_{j\sigma^{\prime}}; ~~~~~~
    \ham_{\text{B}} = -B\summ{i}(n_{i\upa}-n_{i\dna}) 
                = -B\summ{i}(c_{i\upa}^{\dagger}c^{}_{i\upa} - c_{i\dna}^{\dagger}c^{}_{i\dna}).
\label{hamiltonian_total_elaborated}
\end{align}

\noindent
In the above,
$c_{i\sigma}^{\dagger}$($c_{i\sigma}$) creates (annihilates) an electron with spin $\sigma$ at $i^{\text{th}}$ site of the lattice. The sum over $\la ij\ra$ represents the sum over the nearest-neighbor sites of the square lattice. The operator $n_{i\sigma}$ is the electron occupation number operator at $i^{\text{th}}$ site for $\sigma$ spin-projection. The total electron occupation number at $i^{\text{th}}$ site is, therefore, defined as $n_{i} = n_{i\upa} + n_{i\dna}$. The parameter $U$ is the on-site attractive interaction strength, whereas $V$ is the inter-site attractive interaction strength. The $z$-component of the magnetic field is given by $B$, which lies in the lattice plane. $t$ is the usual nearest-neighbor hopping amplitude and finally, $\mu$ is the chemical potential as we work in the grand canonical ensemble.

Looking closely at the interaction terms individually and expanding them in second quantization notation we have:

\begin{align}
    \mcal{H}_{\text{int}}^{\text{onsite}} & = -U\summ{i}n_{i\upa}n_{i\dna} = -U\summ{i}c_{i\upa}^{\dagger}c_{i\upa}
    c_{i\dna}^{\dagger}c_{i\dna}
     = -U\summ{i}c_{i\upa}^{\dagger}c_{i\dna}^{\dagger}
    c_{i\dna}c_{i\upa}
    \label{H_int_onsite_real}
\end{align}

\noindent where we rearranged the Fermionic operators in particle-particle channel or so called pairing channel, using anti-commutation rules. Similarly for the inter-site term:

\begin{align}
    \mcal{H}_{\text{int}}^{\text{nn}} & = -V\summ{\la ij\ra}n_{i}n_{j}
    = -V\summ{i,\delta}n_{i}n_{i+\delta}
     = -V\summ{i,\delta}(n_{i\upa}+n_{i\dna})
    (n_{i+\delta\upa}+n_{i+\delta\dna})
    \notag\\
    &=-V\summ{i,\delta}
    (n_{i\upa}n_{i+\delta\upa}
    +n_{i\dna}n_{i+\delta\dna}
    +n_{i\upa}n_{i+\delta\dna}
    +n_{i\dna}n_{i+\delta\upa}
    )
    \notag\\
    &= \underbrace{-V\summ{i,\delta}c_{i\upa}^{\dagger}c_{i+\delta\upa}^{\dagger}c_{i+\delta\upa}c_{i\upa}}_A~~~ \underbrace{-V\summ{i,\delta}c_{i\dna}^{\dagger}c_{i+\delta\dna}^{\dagger} c_{i+\delta\dna}c_{i\dna}}_B ~~~ \underbrace{-V\summ{i,\delta}c_{i\upa}^{\dagger}c_{i+\delta\dna}^{\dagger} c_{i+\delta\dna}c_{i\upa}}_C ~~~ \underbrace{\quad-V\summ{i,\delta}c_{i+\delta\upa}^{\dagger}c_{i\dna}^{\dagger} c_{i\dna}c_{i+\delta\upa}}_D
    \label{H_int_nn_real}
\end{align}
\noindent
where $\delta$ is an index which denotes unit vectors along the direction of two nearest neighbors, namely $\delta = +\hat{x}$ and $+\hat{y}$. Note that we have grouped terms as $A$,$B$,$C$, and $D$ for the ease of reference. Now terms $A$ and $B$ lead to the possibility of triplet solution with $S_{z}=\pm 1$ (ESP states), while terms $C$ and $D$ lead to singlet and triplet solution with $S_{z}=0$ (OSP states). 

Notice that this model describes the many-particle interaction which in itself is very hard to analyse. We therefore make the mean-field approximation and via the Bogoliubov-de Gennes framework, we solve the Hamiltonian computationally, and present the results in the following sections. The details to the BdG framework are given in section \ref{sec:BdG}.

\subsection{Determination of Relative Phase Angle between different Pairing Correlations}
\label{sec:RelativePhase}

Before we discuss the competition among different superconducting
states, we consider the possibility of obtaining different triplet states from different combinations of pairing mean field parameters.

\paragraph*{Triplet combinations from $\Delta^{\uparrow\downarrow}(\mathbf{k})$:}

\begin{enumerate}
 \item $p_{x}$ state (and correspondingly $p_y$ state):   $\Delta_{x}^{+}=-\Delta_{x}^{-} \in \mathbb{C};~~~~~~\Delta_{y}^{+}=
 \Delta_{y}^{-}=0$
 
 \item $p_{x}\pm ip_{y}$ state :   $\Delta_{x}^{+}=-\Delta_{x}^{-} = C \in \mathbb{C}; ~~~~~
 \Delta_{y}^{+}=-\Delta_{y}^{-}=iC$
\end{enumerate}
In general, $p_{x}$ and $p_{y}$ can be of the form: $p_{x}+e^{i\theta}p_{y}$. With the help of energetics and self-consistency we have checked that $\theta = \frac{\pi}{2}(-\frac{\pi}{2})$ corresponds to the most stable solution, i.e. it takes the form $p_{x}\pm ip_{y}$, independent of the external parameters. The energetics suggests that for all external model parameters, the most stable relative phase angle between $p_x$ and $p_y$ states is $\pi/2$. This allows us to reduce our exploration space $\{ \Delta \}$. Similarly, we fix the phase relation among, $\Delta_{x}^+,\Delta_x^-,\Delta_y^+$ and $\Delta_y^-$.

\paragraph*{Triplet combinations from $\Delta^{\uparrow\uparrow}(\mathbf{k})$ and $\Delta^{\downarrow\downarrow}(\mathbf{k})$ :}

\begin{enumerate}
 \item $\Delta^{\uparrow\uparrow}(k)=2iV\left[\Delta_{x}^{\uparrow}\sin(kx)+
 \Delta_{y}^{\uparrow}\sin(ky)\right]$
 
 \item $\Delta^{\downarrow\downarrow}(k)=2iV\left[\Delta_{x}^{\downarrow}\sin(kx)+
 \Delta_{y}^{\downarrow}\sin(ky)\right]$
\end{enumerate}

\begin{figure}[!t]
    \centering
    \includegraphics[width=\linewidth]{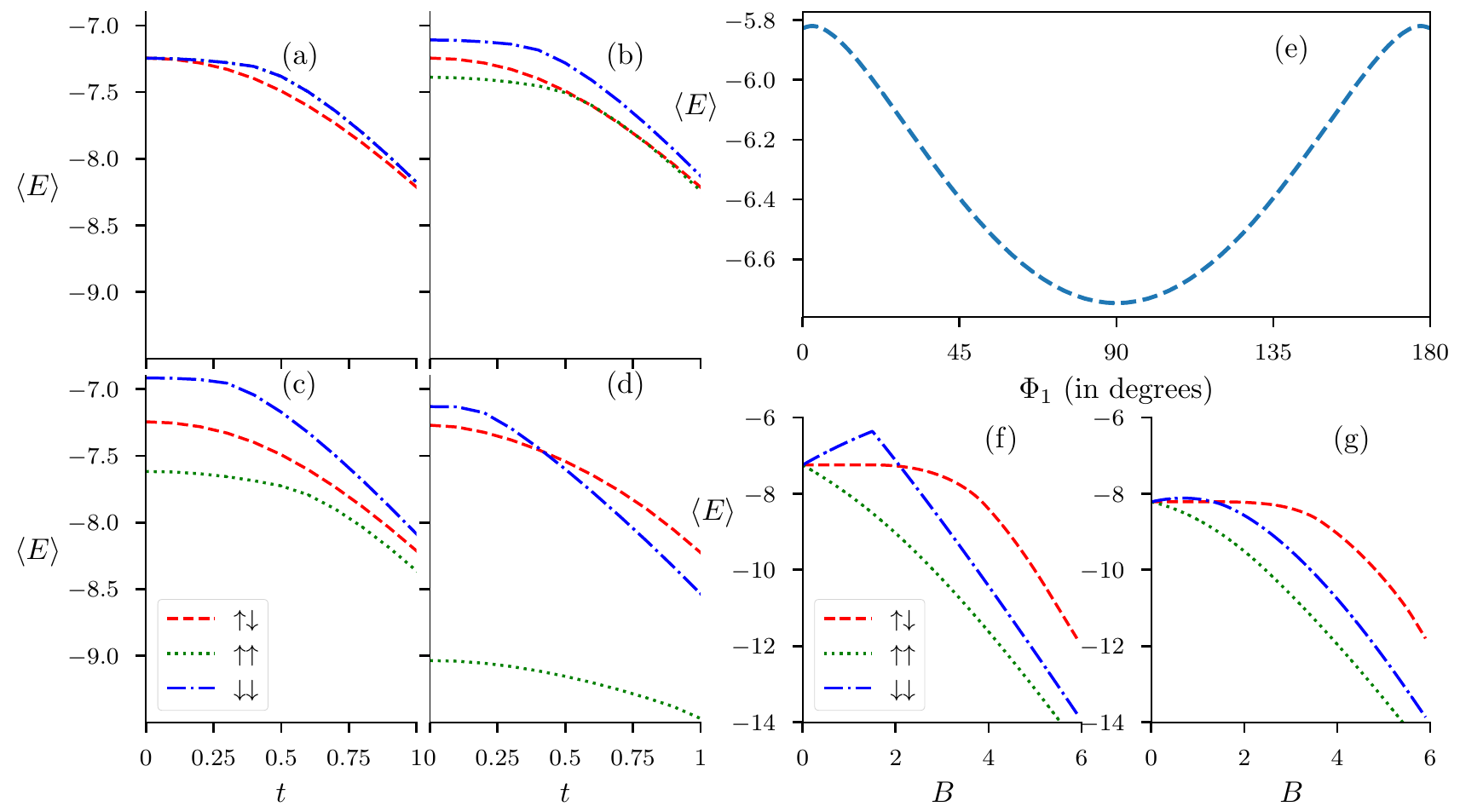}
    \caption{\textbf{(a)-(d).} The plots show the variation of average energy $\la E\ra$ corresponding to
    three triplet states with hopping amplitude $t$ at zero temperature for different magnitudes of the magnetic field, $(a)$ $B=0$, $ (b)$ $B=0.2t$, $(c)$ $B=0.5t$ and 
    $(d)$ $B=2.0t$. \\
	\textbf{(e).} Plot shows the variation of average energy 
    $\langle E\rangle$ with relative phase angle $\Phi_{1}$, where $\Delta_{x}^{\upa} = 0.5,\Delta_{y}^{\upa} = 0.5e^{i\Phi_{1}}, \Delta_{x}^{\dna} = 0.5e^{i\Phi_{2}}$ and $\Delta_{y}^{\dna} = 0.5e^{i(\Phi_1+\Phi_2)}$. The variation shown here is independent of the choice of external parameters like the chemical potential $\mu$, hopping parameter $t$, magnetic field $\mathbf{B}$, and the relative phase between $\Delta_{\delta}^{\uparrow}$ and $\Delta_{\delta}^{\downarrow}$ i.e., $\Phi_2$. This suggests that, independent of the external parameters, the most stable relative phase between $\Delta_{x}^{\sigma}$ and  $\Delta_{y}^{\sigma}$ is $\Phi_1 = \pi/2$.\\
    \textbf{(f)-(g).} The plots shows the variation of average energy $\langle E\rangle$ for the three triplet states with the $z-$component of the magnetic field $B$, for $(f)$ $t=0$ and $(g)$ $t=1$.}
    \label{fig:1}
\end{figure}

Again by comparing energies we find the phase relations between $\Delta_{x}^{\uparrow}$ and $\Delta_{y}^{\uparrow}$ (and correspondingly, between $\Delta_{x}^{\downarrow}$ and $\Delta_{y}^{\downarrow}$) i.e., $\Delta_{y}^{\uparrow} = e^{i\Phi_{1}}\Delta_{x}^{\uparrow}$ (and $\Delta_{y}^{\downarrow} = e^{i\Phi_{1}}\Delta_{x}^{\downarrow}$). We find that
$\Phi_{1} = \frac{\pi}{2}$ minimizes the average energy $\la E\ra$ (Fig. \ref{fig:1} (e)) independent of the external parameters. Note that we also allowed $\uparrow$ and $\downarrow$ parring correlation functions to have a relative phase of $\Phi_2$, and our results are independent of this relative phase. These 
phase relations reduce our parameter space considerably. Though these 
plots (Fig. \ref{fig:1} (e)) are generated on the basis of mere energetics, we also checked
the validity of these relations in self-consistent solutions. Notice that these relations only hold for the correlation functions $\{\Delta_{x}^{\uparrow}, \Delta_{y}^{\uparrow}\}$ (and $\{\Delta_{x}^{\downarrow}, \Delta_{y}^{\downarrow}\}$)  and we cannot, at this stage, comment on the phase difference between $\Delta^{\uparrow\uparrow}(\mathbf{k})$ and $\Delta^{\downarrow\downarrow}(\mathbf{k})$ because these results are independent of $\Phi_2$.

\subsection{Competition among the different triplet states}\label{sec:competetionTripletStates}

In this section we compare the energies of the three triplet states: $|S_{z}=0\rangle$ $\equiv \frac{1}{\sqrt{2}}(|\uparrow\downarrow\rangle + |\downarrow\uparrow\rangle)$, $|S_{z}=1\rangle$ $\equiv|\uparrow\uparrow\rangle$, and $|S_{z}=-1\rangle$ $\equiv|\downarrow\downarrow\rangle$. 
We expect all three states $|S_{z}=0\rangle$, $|S_{z}=1\rangle$, and $|S_{z}=-1\rangle$ to be degenerate in the absence of magnetic field. Furthermore, we expect this degeneracy to be lifted in the presence of magnetic field, and one of the equal spin pairing (ESP) states $|S_{z}=1\rangle$ or $|S_{z}=-1\rangle$ will have lower energy in the presence of an in-plane magnetic field depending on its orientation. In our formulation, singlet and $|S_{z}=0\rangle$ triplet component enter into the Hamiltonian through the superconducting pairing correlation $\Delta^{\uparrow\downarrow}(\mathbf{k})$; and $|S_{z}=1\rangle$, $|S_{z}=-1\rangle$ triplet contributions enter through the triplet superconducting correlations $\Delta^{\uparrow\uparrow}(\mathbf{k})$ and $\Delta^{\downarrow\downarrow}(\mathbf{k})$. Therefore, we expect this behaviour to translate to the superconducting pairing correlations as well.


The analysis is summarised in Fig. \ref{fig:1} (a)-(d) and Fig \ref{fig:1} (f)-(g). In Fig. \ref{fig:1} (a)-(d), we plot the variation of average system energy corresponding to all possible triplet states  (both OSP and ESP) with increasing amplitude of nearest-neighbor hopping parameter $t$. As we focus on the energies of the triplet states, we pick a specific point in the parameter space ($U=0$, $V=1.8$, $\mu = -1.5$) where we found pure-triplet ($p_{x}\pm ip_{y}$ type) state being stabilized in our previous study \cite{nayak2018}. As the interaction strength $V$ is kept fixed while increasing nearest-neighbor hopping amplitude $t$, effectively we are moving from a strong-coupling limit to a weak-coupling limit.


In the strong-coupling limit, where the electron-pairs are tightly bound, we can appropriately apply the well-known physics of spin-singlet and spin-triplet pairing to the spin degrees of freedom associated with the electron-pairs. In this limit, absence of magnetic field leads to degenerate states. In fact, absence of magnetic field provides the freedom to transform ESP states into specific OSP states, via a suitable choice of quantization axis due to rotational symmetry. In the presence of a magnetic field this symmetry is broken and therefore the degeneracy is lifted. In the weak-coupling limit, the system gains energy via de-localization of electrons, and characterization of different triplet states in terms of local spin operators is not valid. Note that we use the phase-lock between different pairing correlations obtained in the previous section -- this reduces the exploration parameter space significantly and gives us the freedom to use only magnitudes of different correlation functions. In Fig. \ref{fig:1} ($a$), at $t=0$, we choose these magnitudes in a manner so that all the triplet states, both OSP and ESP, remain degenerate. With the increase of nearest-neighbour hopping amplitude $t$ we note that the OSP state ($\upa\dna$ in figure) becomes energetically favourable. Furthermore, at any finite value of Zeeman-coupling, $\upa\upa$-type ESP state remain energetically favourable as we expected, while the energy of $\dna\dna$-type ESP state in comparison to OSP state behave differently in different coupling regimes. At moderate values of Zeeman-coupling, the system prefers OSP state to $\dna\dna$-type ESP state in both weak and strong coupling limits (Fig.\ref{fig:1} ($b$)-($c$)). However, at a larger value of Zeeman-coupling, OSP state is completely disfavoured in the weak-coupling limit (Fig.\ref{fig:1} ($d$)). One should take a note of the fact that this particular observation is validated by self-consistent solutions presented in the forthcoming sections. 

Fig. \ref{fig:1} (f)-(g) completes this analysis, as it captures the variation of energy of the triplet states with increasing values of Zeeman-coupling in the two limiting cases ($t = 0$ and $t=1$). At strong-coupling ($t=0$) limit, OSP state is of lower energy compared to $\dna\dna$-type ESP state below a critical Zeeman-coupling ($B\approx 2$). Beyond such critical value of Zeeman-coupling OSP state becomes completely disfavoured energetically with respect to the ESP states. On the other hand, at weak-coupling limit ($t=1$), energy of OSP state remains comparable with $\dna\dna$-type ESP state at moderate values of Zeeman-coupling, and eventually becomes energetically disfavoured at higher values of Zeeman-coupling. On one hand this section provides us with the idea of how OSP and ESP states behave at two different coupling regimes. On the other hand, it provides us an insight into the nature of triplet states and which type is likely to be stabilized at different values of Zeeman-coupling.

\subsection{Ground State Phase Diagrams}\label{sec:GroundStatePD}

In the previous work\cite{nayak2018} we explored the possibility of stabilizing OSP superconducting solutions in the framework of Extended attractive Hubbard model defined on a square lattice in the absence of a magnetic field. The results were based on the set of order parameters \{$\Delta$\} = \{ $\Delta_{s},\Delta_{s^{\ast}},\Delta_{d_{x^2-y^2}}
\Delta_{p_{x}},\Delta_{p_{y}}$ \}, which were obtained using the standard definition of superconducting OPs, defined in literature, from the original set of pairing correlations \{ $\Delta, \Delta_{\delta}^{+}, \Delta_{\delta}^{-}$ \}. Note that $\delta$ is an index for the unit vectors $\delta = \hat{x},\hat{y}$ along two independent directions on a square lattice. So \{ $\Delta, \Delta_{\delta}^{+}, \Delta_{\delta}^{-}$ \} is actually a shorthand notation for \{ $\Delta, \Delta_{x}^{+}, \Delta_{y}^{+}, \Delta_{x}^{-}, \Delta_{y}^{-}$ \}.

In this article, we also include the possibility of the ESP-type superconducting solutions along with OSP-type solutions. Thus our new set of pairing correlations include four more quantities, namely $\Delta_{x}^{\upa}$, $\Delta_{y}^{\upa}$, $\Delta_{x}^{\dna}$, and $\Delta_{y}^{\dna}$. So, our new set of order parameters becomes
\begin{align}
 \{ \Delta \} = \overbrace{
                \{ \Delta,\quad
                \Delta_{\delta}^{+},\quad\Delta_{\delta}^{-}
                }^{\text{OSP}},\quad
                \overbrace{
                \Delta_{\delta}^{\upa},\quad\Delta_{\delta}^{\dna} 
                }^{\text{ESP}}
                \},
\label{Ch4_Eq_new_OP_set}
\end{align}

\noindent
where, again we have used a shorthand notation as explained above. Note that, in Eq.(\ref{Ch4_Eq_new_OP_set}), we have grouped the pairing correlations in two sets, namely OSP and ESP. The OSP group contains pairing correlations ($\Delta$,$\Delta_{\delta}^{+}$,$\Delta_{\delta}^{-}$) that allow the OSP-type triplet states to exist, while the ESP group contains pairing correlations ($\Delta_{\delta}^{\upa}$,$\Delta_{\delta}^{\dna}$) that give rise to ESP-type triplet states. We follow an unrestricted approach, as adopted in \cite{nayak2018}, where OSP group of pairing correlations are allowed to take such forms where they can either make the singlet or the OSP-type triplet, or a mixture of them them to be stabilized in different parameter regimes. On the other hand, ESP group of pairing correlations allow only ESP-type triplet states to be stabilized because of their symmetrized-spin wavefunction (and therefore the orbital part of the wavefunction which is the superconducting gap function, should be anti-symmetric) by definition.

Having defined our order parameters in the above manner, we set out to study the variation of magnitudes of these order parameters with varying chemical potential $\mu$. Average electron density per site $\la n\ra$ is calculated for each value of $\mu$. This allows one to plot magnitudes of the order parameters, defined in Eq.(\ref{Ch4_Eq_new_OP_set}), with varying $\la n\ra$. 
In the self-consistent approach, if we start with a random initial configuration of $\{ \Delta\}$ in the absence of magnetic field, we find individual magnitudes of the converged solutions to be arbitrarily different for different initial runs due to degeneracy. This prevents us to plot a smooth variation of the OPs with $\la n\ra$. In the absence of Zeeman coupling this is quite expected. As in the absence of Zeeman coupling, there is no fixed quantization axis present in the system, possible OSP and ESP-type triplet states can form different linear combinations of corresponding pairing correlations, leaving us with a degenerate set of solutions for each value of $\mu$.  In this case the $d$-vector formalism, defined in the section \ref{sec:d-vec}, comes to our rescue. We find magnitudes of $d_{0}(\vk)$ and $\mbf{d}(\vk)$, averaged over the whole Brillouin zone, to be the best suited order parameters for the singlet and the triplet states respectively, in this scenario. Thus we define our singlet and triplet order parameters in the following way:

\begin{align}
  \Delta_{\text{SP}} = \frac{1}{N_{s}}\summ{\vk} \lv d_{0}(\vk)\rv 
  \quad\text{and}\quad
  \Delta_{\text{TP}} = \frac{1}{N_{s}}\summ{\vk} \lv\mbf{d}(\vk)\rv 
  \label{Ch4_Eq_singlet_triplet_OP_dfn}
\end{align}

\begin{figure}[t!]
    \centering
    \includegraphics[width=\columnwidth]{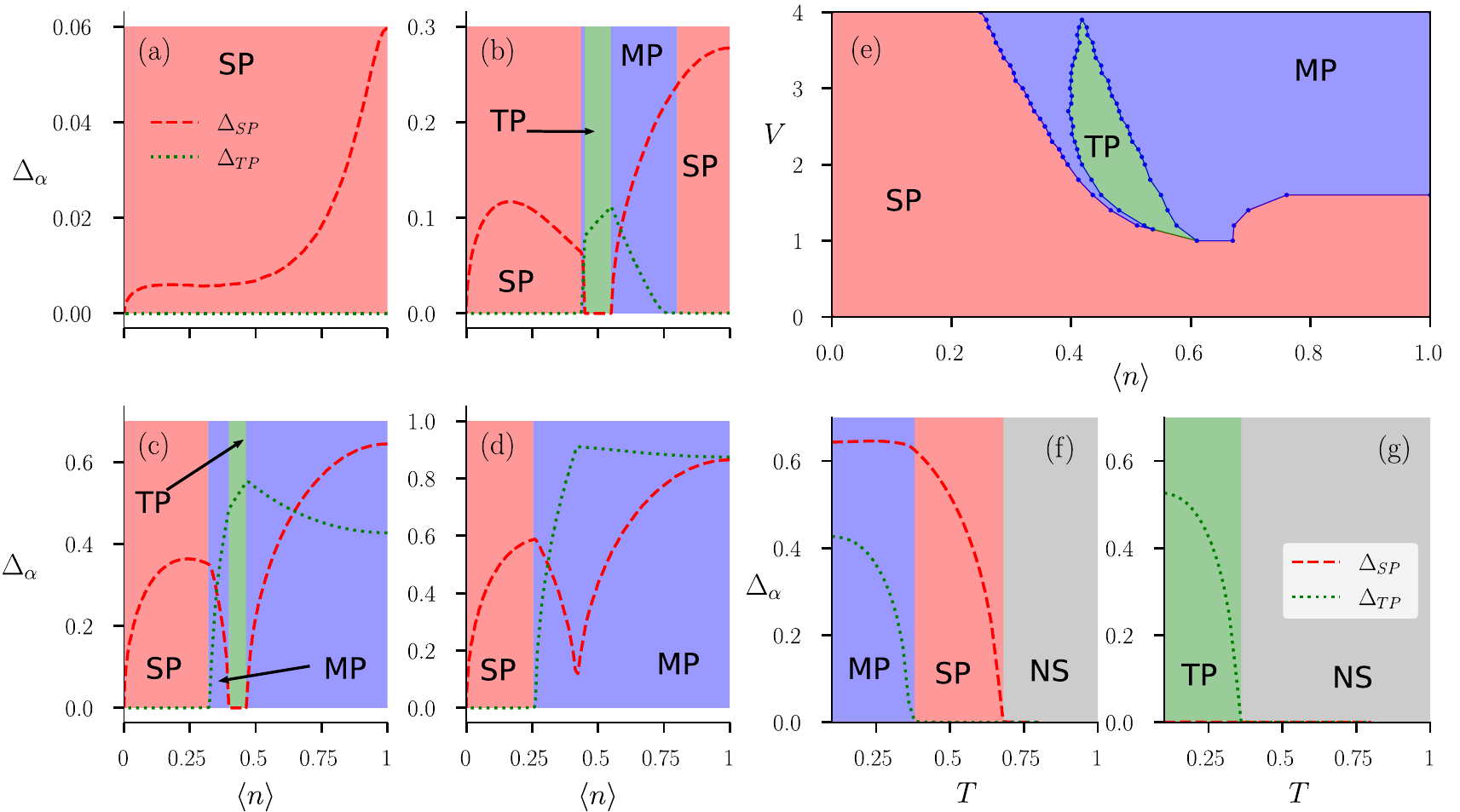}
    \caption{The color in the background indicates the nature of the SC order as,
    red: pure-singlet (SP), green: pure-triplet (TP), and blue: singlet-triplet mixture (MP). The gray 
   background represents non-superconducting region (NS).    \\
    \textbf{(a)-(d).} Variation of singlet $\Delta_{SP}$ and 
    triplet $\Delta_{TP}$ order parameter magnitudes with electron
    density $\la n\ra$ at zero temperature for onsite interaction $U=1$ 
    and inter-site interaction strength: 
    $(a)$ $V = 0.4t$, $(b)$ $V = 1.6t$, $(c)$ $V = 3.0t$ and $(d)$ $V = 4.0t$.\\
    \textbf{(e).} Plot (e) shows the $V$-$n$ ground state phase diagram at $B=0$, $U=t$. \\
    \textbf{(f)-(g).} Variation of singlet $\Delta_{SP}$ and triplet
    $\Delta_{TP}$ order parameter magnitudes with temperature $T$ in the 
    absence of Zeeman-coupling for on-site interaction $U=t$, 
    inter-site interaction $V=3t$, and average electron density:
    ($f$) $\la n\ra=1.0$ and ($g$) $\la n\ra=0.44$.
    }
    \label{fig:2}
\end{figure}

Using the newly defined singlet and triplet order parameters in Eq.(\ref{Ch4_Eq_singlet_triplet_OP_dfn}), we study the variation of these order parameters with average electron density per site $\la n\ra$. In Fig. \ref{fig:2} (a)-(d), we plot the variation of singlet order parameter $\Delta_{SP}$  and triplet order parameter $\Delta_{TP}$ with $\la n\ra$ for different values of  inter-site attraction $V$. We keep on-site attraction at a fixed value ($U=t$), as it is the inter-site attraction $V$ that plays a role in stabilizing different unconventional superconducting orders as deduced from our previous works\cite{nayak2018}. When inter-site attraction $V$ is small ($V=0.4t$) compared to on-site attraction $U$, it is the singlet phase that occupies the whole density region (Fig. \ref{fig:2} ($a$)). At $V=1.6t$, singlet phase occurs both at lower density region and near half-filling (Fig. \ref{fig:2} ($b$)). Though orbital information of the order parameter is averaged out in defining the singlet and triplet order parameters, it's easy to suggest that the singlet phase near half-filling is $d_{x^2-y^2}$ type, while the singlet phase near low-density region is of $s+s^{\ast}$ type based off our previous work\cite{nayak2018}. Pure triplet phase occurs near quarter-filling, while singlet-triplet mixed parity phase occurs between pure triplet phase and singlet phase of $d_{x^2-y^2}$ type. The pure triplet phase consists of all three types of triplet pairings due to absence of magnetic field and therefore it is meaningless to give it a specific name. It is important to note that pure-$d_{x^2-y^2}$ type singlet phase occurs at low-$V$ region, when $V$ is still larger than on-site interaction $U$. At higher value of inter-site attraction ($V=3t$), pure-$d_{x^2-y^2}$ type singlet phase vanishes (Fig. \ref{fig:2} ($c$)), while singlet-triplet mixed phase occupies the corresponding density region (near half-filling). At $V=3t$ a small window of singlet-triplet mixed phase occurs between $s+s^{\ast}$ type singlet and pure triplet phase. If we increase $V$ further ($V=4t$), most of the density region is occupied by dominant singlet-triplet mixed phase, while $s+s^{\ast}$ type singlet phase still occurs at low density region (Fig. \ref{fig:2} ($d$)).


This variation of order parameters with average electron density motivates us to draw a $V$-$\la n\ra$ ground state phase diagram in the absence of a magnetic field (Fig. \ref{fig:2} (e)). We notice that when the inter-site interaction $V$ is less than the on-site interaction, i.e. $V < t$, the whole density region is occupied by the pure singlet phase (red in color). This is expected as a dominant on-site attraction is known to stabilize singlet $s$-wave SC order. While $U$ is stabilizing singlet $s$-wave SC order, $V$ is also playing an important role in stabilizing the singlet $s^{\ast}$ and the unconventional $d_{x^2-y^2}$ SC orders in this region along with $s$-wave. When the lower density region prefers singlet $s$ and $s^{\ast}$ SC orders to other phases, near half-filling it is the singlet $d_{x^2-y^2}$-wave that prevails. Notice that these inferences come from a combination of the present work and the previous work \cite{nayak2018}. As inter-site interaction $V$ becomes larger compared to on-site interaction $U$, it not only stabilizes singlet $s^{\ast}$ and $d_{x^2-y^2}$ SC orders, it also helps pure triplet and mixed-parity phases to get stabilized at different density regimes. The pure triplet phase (green in color) occurs in a delta-shaped region near quarter filling ($0.4 \leq\la n\ra \leq 0.6$), when the inter-site attraction $V$ is higher than the on-site attraction $U$ but not as large as $V=4t$. With $V$ getting larger, the mixed-parity (blue in color) phase becomes more stable, specially in the higher density region. It's interesting to note that $d_{x^2-y^2}$-type singlet phase is prone to get stabilized near half-filling region and pure triplet phase has a tendency to get stabilized near quarter-filling. But depending on the strength of inter-site interaction the system gains energy by stabilizing a new mixed-parity (possibly $d+p$-type extrapolating from our previous study) state, near the higher density region. It's also important to note that, in the absence of a magnetic field, the phase diagram is very similar to our previous study\cite{nayak2018}. The only difference is that in the triplet region, we expect the phase to be in a superposition of all possible triplet orders, which was not possible earlier because of the limitations of our exploration space. In the section where we study the effects of magnetic field, Sec. \ref{sec:ZeemanCoupling}, the phase diagram will change drastically since now we have the possibility of a competition between different triplet phases.

\subsection{Finite Temperature Study}\label{sec:FiniteTempStudy}


After exploring the stability of singlet, triplet, and mixed-parity states in the ground state of the extended attractive Hubbard model on a square lattice, we intend to study the behavior of the corresponding SC states
at finite temperatures. General consensus is that thermal excitations destroy superconductivity. So the expectation is that the magnitudes of the superconducting order parameters, defined in our system, will decrease and eventually vanish when temperature is increased beyond a critical value. It is well-known that beyond this critical temperature $T_{c}$ superconductors make a transition from superconducting state to the normal state. As we limit ourselves to the framework of mean-field theory, it is not our aim to infer about the specific values of $T_{c}$ we obtain in our calculations. We rather intend to study how different SC orders react to the onset of finite temperature.

\begin{figure}[!t]
    \centering
    \includegraphics[width=\columnwidth]{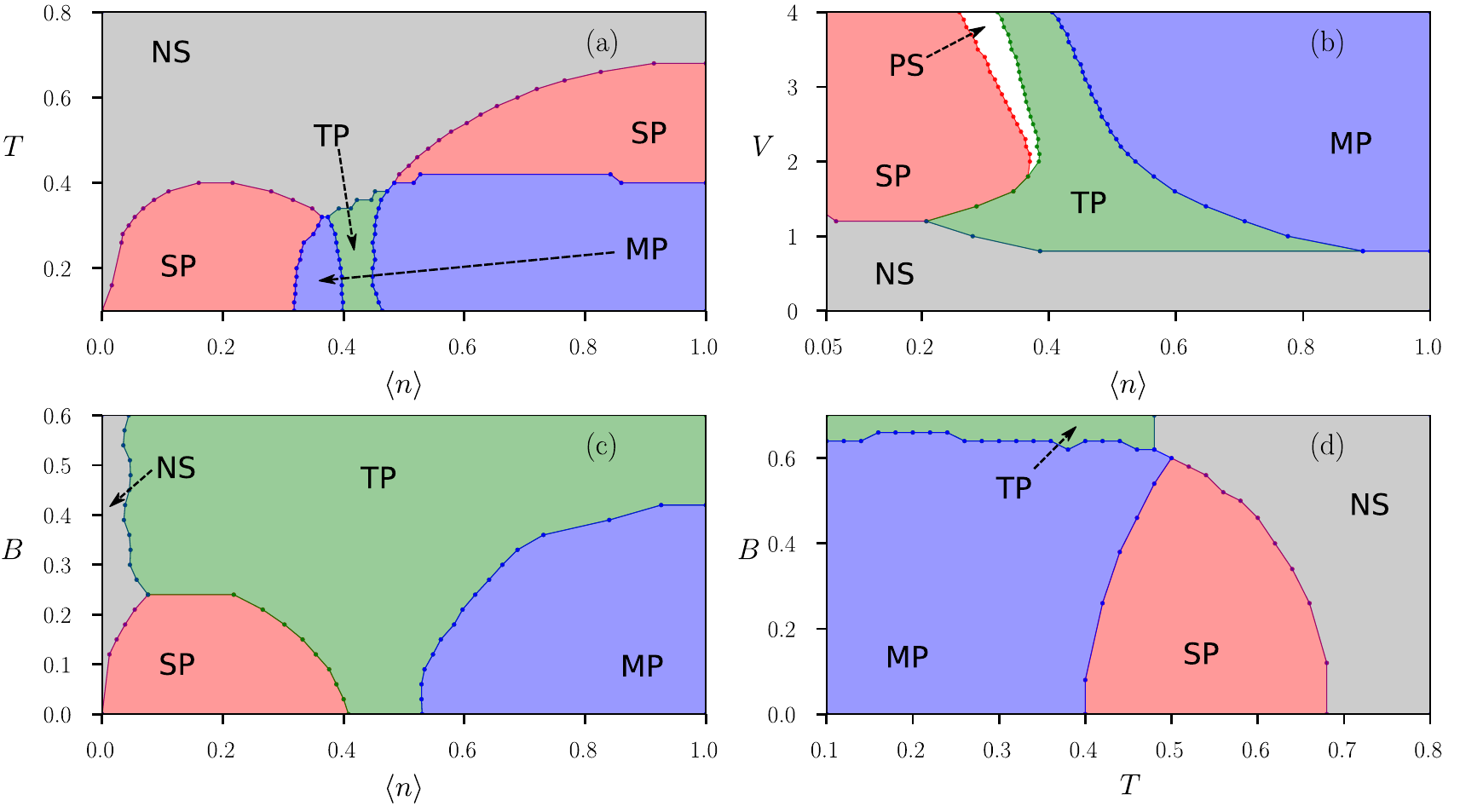}
    \caption{The color in the background indicates the nature of the SC order as,
    red: pure-singlet (SP), green: pure-triplet (TP), blue: singlet-triplet mixture (MP), and gray: non-superconducting region (NS). White region represents phase separation (PS). \\
    \textbf{(a).} $T$-$\la n\ra$ ground state phase diagram, in the absence
    of a magnetic field, at $U=t$ and $V=3t$. \\
    \textbf{(b).} $V$-$\la n\ra$ ground state ground state phase diagram at finite magnetic field, $B=0.1t$ 
    and $U=t$.\\
    \textbf{(c).} $B$-$\la n\ra$ ground state phase diagram at zero temperature, $V=2t$, $U=t$. \\
    \textbf{(d).} $B$-$T$ phase diagram at $U=t$ and $V=3t$. }
    \label{fig:3}
\end{figure}

In Fig. \ref{fig:2} (f)-(g), we plot the variation of singlet $\Delta_{SP}$ and triplet $\Delta_{TP}$ order parameter with temperature for two different points in parameter space taken from the ground state $V$-$\la n\ra$ phase diagram. In Fig. \ref{fig:2} ($f$), we start with a mixed-parity state exactly at half-filling $\la n\ra=1$. While the red dashed line represents the singlet order parameter $\Delta_{SP}$, the green dotted line represents triplet order parameter $\Delta_{TP}$. As temperature is increased, the triplet order parameter starts to fall off and around $T\approx 0.38$ it vanishes completely. Most interestingly, the singlet order parameter $\Delta_{SP}$ remains almost constant while $\Delta_{TP}$ falls off. It seems like the $d_{x^2-y^2}$ type singlet remains protected as long as the triplet order parameter is finite. At $T\approx 0.38$ the system makes a transition from a mixed-parity SC state to a pure $d_{x^2-y^2}$ type singlet superconducting state. If temperature is increased further $\Delta_{SP}$ starts to fall off and it completely vanishes at $T\approx 0.68$. Thus Fig. \ref{fig:2} ($f$) depicts how the system makes a transition from a mixed-parity superconducting state to a pure singlet superconducting state before making a transition to a non-superconducting state. In Fig. \ref{fig:2} ($g$), we start with a different point of the parameter space, again taken from the ground state $V$-$\la n\ra$ phase diagram. In this case our zero temperature initial choice is at density $\la n\ra = 0.44$, where we find pure triplet state being stable. Notably with the increase of temperature the triplet order parameter $\Delta_{TP}$ falls off exactly the way it did for the previous case and the temperature where $\Delta_{TP}$ completely vanishes is exactly same as the previous case, i.e., $T\approx 0.38$. But in this case the triplet SC state makes a direct transition to non-superconducting state. 

Finally we summarize these results in Fig. \ref{fig:3} (a), where we present the $T$-$\la n\ra$ phase diagram to complete the analysis of the finite temperature effects. In the $T$-$\la n\ra$ phase diagram we find that every superconducting phases at zero temperature makes a transition from superconducting to non-superconducting phase when temperature is increased. While the transition temperature for singlet phase takes a dome-shaped feature as a function of density, for triplet and mixed-parity states the transition temperature does not have such sharp feature with varying density. As discussed earlier, we find four  different sectors of such transition. In three sectors the system makes a direct transition from superconducting state to non-superconducting state retaining the specific form of pairing symmetry, while the mixed-parity state in the density region $0.5\leq \la n\ra \leq 1.0$ makes a transition to singlet superconducting state before making a transition to non-superconducting state.

\subsection{Effect of Zeeman Coupling}
\label{sec:ZeemanCoupling}



After analysing the behaviour of unconventional superconducting phases at zero and finite temperature in the absence of a magnetic field, we study the effects of the magnetic field on these phases in different regions of the parameter space. We limit our study to the effect of Zeeman-coupling to an external magnetic field. Thus, we ignore the effects of an external magnetic field on orbital degrees of freedom and confine ourselves to investigate the effects of the magnetic field on the spin degrees of freedom only. An in-plane magnetic field serves our purpose. Motivated by the zero temperature $V$-$\la n\ra$ phase diagram presented in Fig. \ref{fig:2} (e), we take interest in knowing how this phase diagram changes with the onset of a Zeeman field. Figure Fig. \ref{fig:3} (b) shows the $V-\la n\ra$ phase diagram in the the presence of magnetic field with $B = 0.1t$. The first thing we notice is the appearance of a non-superconducting region (gray in color), which is quite expected as the magnetic field destroys superconductivity. It is interesting to note that the non-superconducting region appears at lower values of $V$ suggesting that for the superconductivity to survive it really is a competition between the inter-site interaction strength and the magnetic field strength.

\begin{figure}[!t]
    \centering
    \includegraphics[width=10cm]{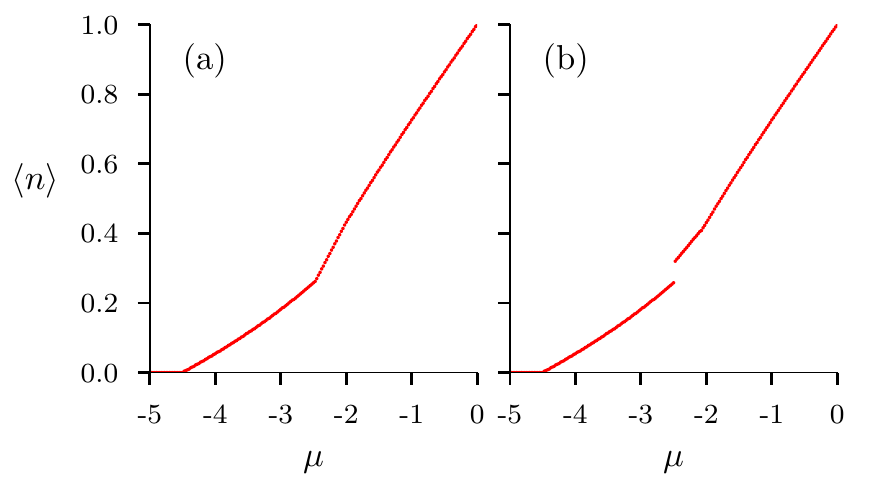}
    \caption{Variation of average electron density $\la n\ra$
    with chemical potential $\mu$ for, $(a)$ $B=0$ and $B=0.1t$.
    The discontinuity in $(b)$ corresponds to a region of phase separation.}
    \label{fig:4}
\end{figure}

From the ground state $V$-$\la n\ra$ phase diagram in the absence of a magnetic field Fig. \ref{fig:2} (e), we notice that when inter-site interaction $V$ is less than the on-site interaction $U$, the whole density region is occupied by singlet superconducting phase. When the Zeeman field is switched on and the strength of the Zeeman field is maintained at $B = 0.1t$, it turns out to be strong enough to destroy superconductivity when $V$ is approximately less than $0.8t$. So any superconducting phase occurring below $V\approx 0.8t$ is destroyed by a Zeeman field of strength $B = 0.1t$. In the absence of a magnetic field, there is a pure  $d_{x^2-y^2}$ type singlet phase, which is stable near half-filling. It appears that presence of an external magnetic field of strength $B = 0.1t$, destroys such pure $d_{x^2-y^2}$ type superconducting phase, as for any strength of $V$ (within the scale provided in Fig. \ref{fig:3} (b)), only the singlet-triplet mixed state gets stabilized. Most interestingly, the region, where pure triplet superconducting phase is stabilized in $V$-$\la n\ra$ phase diagram, gets larger in presence of the Zeeman-coupling suggesting that the pure triplet superconducting phase gets enhanced by the onset of a magnetic field, this is due to the allowed ESP superconducting correlations.


Another important aspect we notice is that there is a region of phase separation appearing between the pure singlet phase and the pure triplet phase when $V$ is approximately greater than $2t$. To look deeper and confirm this aspect of phase separation, we study the variation of average electron density per site $\la n\ra$ with the chemical potential $\mu$ at two values of Zeeman-coupling, ($a$) $B=0$, and ($b$) $B=0.1t$ presented in Fig. \ref{fig:4}. For the purpose of the above-mentioned illustration we keep the on-site interaction at $U=t$, while the inter-site interaction is kept fixed at $V=4t$. While in the absence of a Zeeman-coupling we see a continuous variation of $\la n\ra$ with $\mu$, in the presence of a Zeeman-coupling, we clearly notice a discontinuity in the electron density at $\mu\approx -2.5$. The discontinuity in the density is of the order $\delta_{\la n\ra}\approx 0.1$, at $V=4t$ (Fig. \ref{fig:3} (b)).


To completely understand the effect of Zeeman-coupling on the superconducting phases, we present, in Fig. \ref{fig:3} (c), a zero temperature $B$-$\la n\ra$ phase diagram at $U=t$ and $V=2t$. At $B=0$, $U=t$ and $V=2t$, singlet, triplet, and mixed-parity superconducting phases occupy different sectors of density regions, as observed in the ground state $V$-$\la n\ra$ phase diagram in the absence of the Zeeman-coupling to an external magnetic field. As the strength of the magnetic field is increased it appears that the pure triplet superconducting state becomes favourable. In fact, the pure singlet superconducting phase disappears from the whole density region when the strength of the magnetic field becomes approximately greater than $0.23t$, while the mixed-parity state completely disappears from the whole density region when $B$ becomes larger than $\sim 0.4t$. Eventually at $B>0.4t$ only pure triplet superconducting phase survives throughout the whole density region. Though we must take a note of the fact that a non-superconducting region (gray in color) prevails over the other phases at lower electron densities. The information we gather from Fig. \ref{fig:3} (c) is of two-fold nature. Firstly, it is clear that Zeeman-coupling to an external magnetic field at low strength favours pure triplet superconductivity. Secondly, the disappearance of pure singlet and mixed-parity state indicates the possibility of disappearance of OSP type superconducting states and appearance of ESP type superconducting states beyond certain strength of the external magnetic field. As OSP type states give rise to singlet states like $s$,$s^{\ast}$ and $d_{x^2-y^2}$ etc., and ESP type states give rise to pure triplet superconducting phases, the above-mentioned behavior is well-explained. 

We conclude this section by looking at the effect of Zeeman-coupling at finite temperatures. Fig. \ref{fig:3} (d) shows the $B$-$T$ phase diagram to describe these effects at a specific point of the parameter space, namely $U=1$, $V=3$ and $\mu=0$. As expected, we see that for high enough temperatures, the normal state takes over for all kinds of superconducting phases. Also notice that at higher magnetic field strength, the superconductivity is destroyed at lower critical temperatures, in contrast to the critical temperatures at lower magnetic field strengths.

\subsection{Characterization of Phases}\label{sec:Characterization}

\begin{figure}[!t]
  \begin{center}
    \includegraphics[width=\columnwidth]{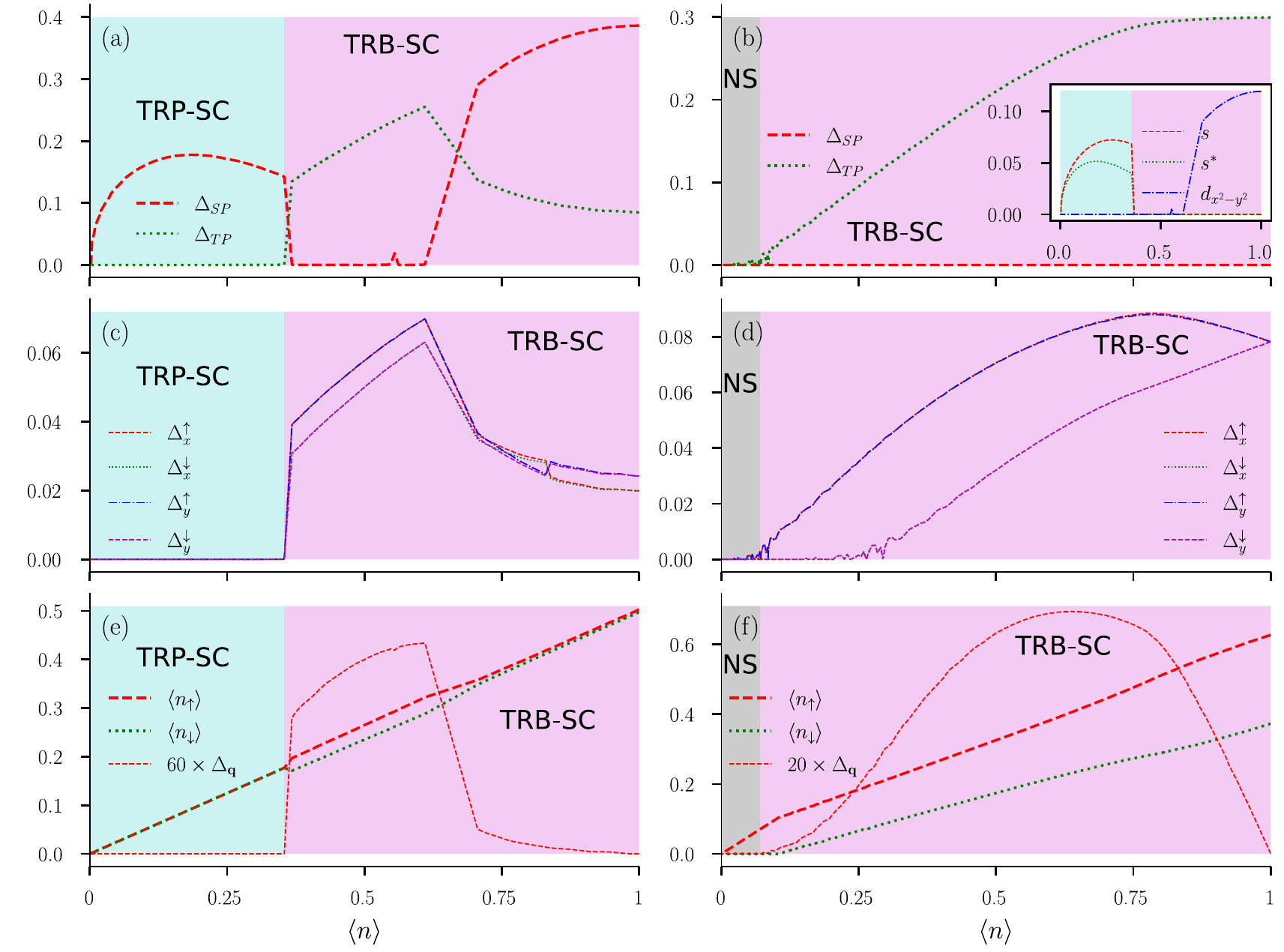}
  \end{center}
  \caption{The color in the background indicates the nature of the
  SC order as, cyan: time reversal symmetry preserving (TRP) SC order, magenta: 
  time reversal symmetry breaking (TRB) SC order, 
  and gray: non-superconducting (NS) region.\\
	{\bf (a)-(b).} Variation of $\Delta_{SP}$ and $\Delta_{TP}$
	as a function of average electron density $\la n\ra$ for
	(a) $B = 0.12t$, and (b) $B = 0.6t$. Inset in (b) displays orbital part of the 
	singlet component of SC order plotted in panel (a).
	Note that the presence of $s$-wave and $s^{\ast}$-wave in
	pure-singlet phase, and the presence of 
	$d_{x^{2}-y^{2}}$-wave character in the mixed-parity phase
	is confirmed from this plot.	 \newline
	{\bf (c)-(d).} SC pair correlation averages 
	$\Delta_{x}^{\upa}$, $\Delta_{x}^{\dna}$, $\Delta_{y}^{\upa}$, 
	and	$\Delta_{y}^{\dna}$ as a function of $\la n\ra$ for 
	(c) $B = 0.12t$, and (d) $B = 0.6t$.\newline
	{\bf (e)-(f).} Variation of spin-resolved average electron
	densities ($\la n_{\upa}\ra$ and $\la n_{\dna}\ra$)
	and $\Delta_{\mbf{q}}$, representing 
	the non-unitarity of the SC order for 
	(e) $B = 0.12t$, and (f) $B = 0.6t$.\\
	For clarity $\Delta_{\mbf{q}}$ is scaled by appropriate factors in 
	panels (e) and (f). We have used $U = t$ and $V = 2t$ for all the data shown in this figure.
	}
  \label{fig:5}
\end{figure}

\begin{figure}[!t]
    \centering
    \includegraphics[width=0.9\columnwidth]{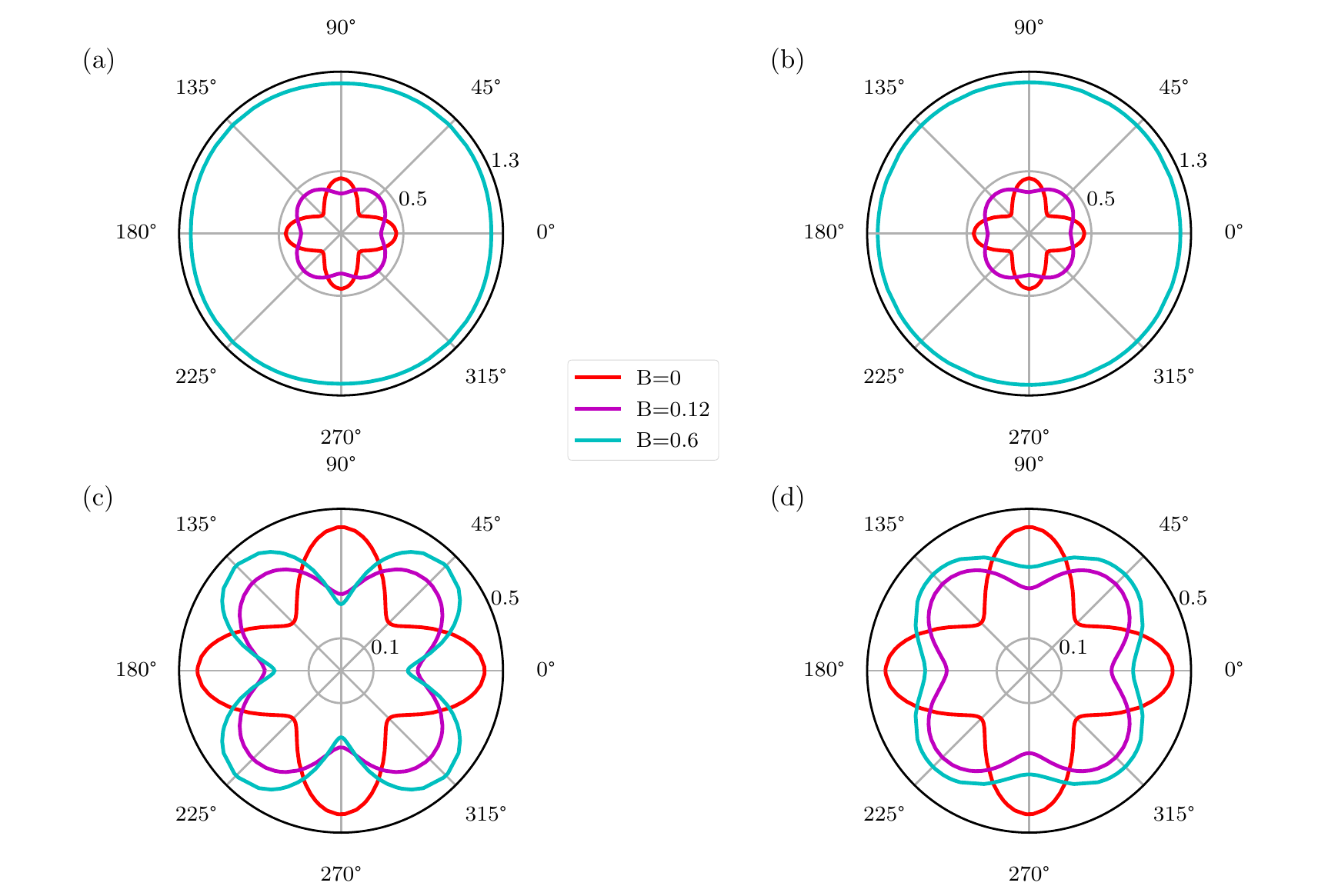}
    \caption{Angular variation of the superconducting energy gaps corresponding to the two pseudo-spin degrees of freedom, for different values of Zeeman field strength $B = 0$, $B = 0.12t$, and $B = 0.6t$. The values of other parameters are fixed as $U = t$, $V = 2t$, $\langle n\rangle \approx 0.6$ for all the panels in this figure.\newline
    {\bf (a)-(b).} Superconducting energy gaps, corresponding to two pseudo-spins, at the $\upa$-spin Fermi surface in the presence of magnetic field.\newline
    {\bf (c)-(d).} Superconducting energy gaps, corresponding to two pseudo-spins, at the $\dna$-spin Fermi surface in the presence of magnetic field.\newline 
    }
    \label{fig:6}
\end{figure}

Our main aim in this study is to provide an identification, based on explicit energy minimization, of distinct-symmetry superconducting order parameters in the presence of a Zeeman coupling. So far we have classified these superconducting states in terms of opposite spin pairing (OSP) and equal spin pairing (ESP) states, leading to a description of minimum energy solutions in terms of pure singlet, pure triplet, and mixed-parity superconducting states. While, the main classification is done based on the nature of the spin state of the system we do mention different pairing symmetries, such as $s$-wave, $s^{\ast}$-wave, $p_{x}+ip_{y}$-wave and $d_{x^2-y^2}$-wave to define the nature of the orbital part of the gap function \cite{nayak2018,batra2019topological}. In the case of pure singlet, or pure triplet phase the nature of the orbital part can be easily understood, while for mixed-parity states such a separation does not exist. In Fig. \ref{fig:5} we look at various quantities related to the superconducting order parameters at zero temperature with varying average electron density per site $\la n\ra $. Results presented in the first column of panels in Fig. \ref{fig:5} correspond to the case when the Zeeman field $B = 0.12t$. On the other hand, the second column represents data for a relatively higher value of the Zeeman field, $B = 0.6t$. The on-site attraction potential $U$ and the inter-site attraction potential $V$ are kept fixed at $U = t$ and $V = 2t$, for all the plots in Fig.\ref{fig:5}. 

In Fig. \ref{fig:5} ($a$), we plot the singlet and triplet order parameters as defined previously. We observe three different sectors in the whole density profile: pure singlet at lower density region, pure triplet near quarter filling, and mixed-parity solution around half filling. Further characterization of the singlet phases, based on the definitions of $s-$wave, $s*-$wave, and $d$-wave are shown in the inset of Fig. \ref{fig:5} ($b$). Note that near half filling, where the singlet phase shows $d$-type pairing (inset of Fig. \ref{fig:5} (b)), it is a mixed parity state with $d_{x^2-y^2}$  as a singlet component. The presence of $s$-wave and $s^{\ast}$-wave character in pure singlet phase and $d_{x^{2}-y^{2}}$-wave character in the mixed-parity phase is consistent with our previous findings when the Zeeman field is absent\cite{nayak2018}. These pairing symmetries do not change with the addition of magnetic field, however, their relative magnitudes do change as discussed in the previous sections. Finite values of singlet order parameter $\Delta_{SP}$ and triplet order parameter $\Delta_{TP}$ at different sectors of the doping regime confirm the existence of pure-singlet, pure-triplet, and mixed-parity phases in the system at a lower strength of the magnetic field, $B = 0.12t$. The background color indicates the time reversal symmetry of the superconducting phase ($d$-vector formalism also allows us to probe the time reversal symmetry in a superconducting phase, refer section \ref{sec:d-vec}). Notice that even though Fig. \ref{fig:5} ($a$) is generated for non-zero magnetic field $B=0.12t$, we see the existence of time reversal invariant (TRP-SC) superconducting phase appearing for lower densities. This is the enduring $s-$wave superconducting phase which is present in the absence of magnetic field at lower densities (Fig. \ref{fig:3}). We can see from Fig. \ref{fig:5} ($b$) that as we increase the magnetic field $B=0.6t$, the remnant time reversal invariant phase is destroyed. Of course, this transition is smooth with the increase of Zeeman field. We also notice that the mixed parity phase breaks time reversal, in order to take advantage of the external magnetic field. At a very low electron density, superconductivity completely vanishes, giving rise to the non-superconducting (NS) region. Fig. \ref{fig:5} (a) and (b) shows that, in the presence of a magnetic field, it is the ESP-type correlation that becomes more favourable, although at low enough Zeeman fields, other correlations can also survive.

In Fig. \ref{fig:5}. (c) and (d) we plot the magnitudes of the ESP-type correlations, $\{\Delta_{x}^{\upa}, \Delta_{x}^{\dna}, \Delta_{y}^{\upa}, \Delta_{y}^{\dna}\}$, for the cases presented in Fig. \ref{fig:5}. (a) and (b) respectively. we observe that in the pure triplet state, near quarter filling, $\Delta_{x}^{\upa}$ and $\Delta_{y}^{\upa}$ become equal in magnitude. Similarly magnitudes of $\Delta_{x}^{\dna}$ and $\Delta_{y}^{\dna}$ also become equal, while $\Delta_{\delta}^{\upa}$ and $\Delta_{\delta}^{\dna}$ acquire different values due to the Zeeman coupling. This observation was also inferred from section \ref{sec:RelativePhase}. Interestingly, even in the presence of Zeeman field, this relation among the triplet ESP correlations appears to change as the $d_{x^2-y^2}$ singlet component becomes dominant near half filling. The character of $d_{x^2-y^2}$ breaks the symmetry between the $x$ and $y$ components of the correlation functions, and as $d_{x^2-y^2}$ wave state becomes more dominant near half filling, this character also shows up in the ESP pairing correlations, forcing $\Delta_{x}^{\upa}$ and $\Delta_{x}^{\dna}$ to become equal in magnitude and similarly for the $y$ component as well. Now as we increase the magnetic field, the pure triplet phase takes over most of the density profile and naturally, only time reversal breaking character of the ESP pairing correlations remains. There is however, an interesting observation that exactly at half filling the $\uparrow$ and $\downarrow$ symmetry is forced upon the triplet phase, even in the presence of strong Zeeman field. Since this formalism is intrinsically particle-hole symmetric, at half-filling we see this symmetry showing up as the symmetry between $\uparrow$ and $\downarrow$ spins which is why all the ESP correlations functions, even in the presence of Zeeman field are forced to take the same values.

In Fig. \ref{fig:5} ($e$) spin-resolved average electron densities are plotted along with the magnitude of the ${\bf q}$ vector averaged over the Brillouin zone, $\Delta_{\bf q} \equiv \sum_{\bf k} |{\bf q}|$. Here ${\bf q}-$vector is defined in terms of the ${\bf d}-$vector as ${\bf q}=i~{\bf d}\times {\bf d}^*$. The ${\bf q}({\bf k})-$vector \cite{mackenzie2003superconductivity} physically represents the net spin average present in the pairing state with momentum $\vk$. This does not always ensure a net total spin moment, averaged over the Fermi surface. However, it entails the fact that the pair correlation for $\upa$-spin electrons is different than that of the $\dna$-spin electrons, thus non-unitarity of a SC order is usually associated with the breaking of time reversal symmetry (TRS)\cite{sigrist1991phenomenological}. Non-unitarity also implies the opening of two distinct SC energy gaps, driven by the breaking of TRS. In the pure triplet region we observe the possibility of the formation of an effective magnetic moment, however, interestingly enough, even in the presence of (small) magnetic field, in the regions where singlet superconducting phases are present, the spin-resolved electron densities are always equal and thus the magnetic moment is always zero.  The magnitude of the ${\bf q}$ vector, $\Delta_{\bf q}$ tells us that the gap function is non-unitary (unitary when $\Delta_{\bf q}=0$) in nature when the triplet order parameter is finite. Note that non-unitarity of triplet gap function is defined in section \ref{sec:d-vec}. From Fig. \ref{fig:5} (e) and (f), we see that non-unitarity is enhanced due to Zeeman field which is in correspondence to the enhancement of the pure triplet phase. It is again interesting to note that quarter filling seems to be the most favourable region for a non-unitary phase to exist and as we go close to half-filling, the non-unitarity is destroyed, consequently, the triplet superconducting phase becomes unitary exactly at half-filling due to the symmetry of the Hamiltonian.

We conclude the section by presenting, in Fig. \ref{fig:6}, the angular variation of the superconducting energy gaps corresponding to the two pseudo-spin degrees of freedom, in the presence of Zeeman field. We choose a representative point at $U = t$, $V = 2t$, and $\la n\ra\approx 0.6$. In the presence of the Zeeman field, the non-interacting Fermi surface splits into two, resulting in separate Fermi surfaces for $\upa$-spin and $\dna$-spin electrons. Interactions lead to a gap opening at these Fermi surfaces, resulting in two gaps as the solution of a $4 \times 4$ matrix at a given ${\bf k}_F$-point leads to two pairs of $\pm E({\bf k_F})$ eigenvalues. Although it may not be straightforward to measure these two gaps in experiments, we nevertheless provide the description in terms of two distinct gaps. In Fig. \ref{fig:6} (a) and (b) we plot the magnitude of these two distinct superconducting energy gaps at $\upa$-spin Fermi surface. Similarly, in Fig. \ref{fig:6} (c) and (d) we plot the magnitude of gaps at $\dna$-spin Fermi surface. This angular variation of the superconducting energy gaps encodes the symmetry of the underlying superconducting state: while at $B = 0$, a $d_{x^2-y^2}$-wave character is prominent from the symmetry of the gap, at $B = 0.12t$ the chiral $p_{x}\pm i p_{y}$ seems to be the dominant symmetry present in the gap structure. Interestingly, at a higher value of the magnetic field, $B = 0.6t$, one of the SC gaps approaches to an isotropic form while the other one retains the $p_{x}\pm i p_{y}$ character.

\section{Conclusions}

Using the unrestricted BdG mean-field approach to the EAHM, we study the competition among different types of superconducting orders. The unique aspect of our study is that we rely on energetics for obtaining the superconducting solutions with different order parameter symmetries.
We find rich ground state phase diagrams with exotic phases such as unitary, non-unitary, pure triplet and mixed parity states. Based on energetics, we provide an understanding of specific phase relations between different superconducting order parameters.
We find an interesting competition between the anti-aligned, w.r.t. the Zeeman field, $\downarrow\downarrow$ ESP state and the OSP state. At moderate values of the Zeeman field, the system prefers the OSP state however, at larger values of Zeeman-coupling, OSP state is completely disfavoured in the weak-coupling limit. The $\uparrow\uparrow$ ESP state is favoured in the presence of Zeeman field. We use the $d$-vector formalism to further distinguish and characterize the different triplet phases.
Pure triplet phase is stable near quarter filling and singlet-triplet mixed phase is favoured near half-filling. The definition of the triplet order parameters being rotationally invariant allows us to study their behaviour even in the presence of magnetic field consistently. Finite temperature study shows transitions from a mixed parity ground state to a pure singlet phase of $d_{x^2-y^2}-$ type, and finally to a non-superconducting phase.
In the presence of Zeeman field, the pure triplet phase dominates the phase diagram, suppressing the pure singlet and mixed parity phases. It is interesting to note that the mixed parity phase is relatively more robust against the Zeeman field as compared to the pure singlet phase. Finally, we discuss the key distinction between different phases in terms of the angular shapes of the gap functions that are experimentally measurable. Our results will be particularly useful in understanding the effect of planar magnetic fields on the nature of superconducting states stabilized in quasi two-dimensional systems, such as atomically thin conducting interfaces between insulating oxides.

\section{Methods}
\label{sec_ModelMethod}

\subsection{Mean-Field Decoupling in Pairing Channel}
\label{subsec_MF_decoupling}

Both the interaction terms, the on-site interaction and the inter-site interaction term, are represented by two-body operator terms in the second quantization formalism. To reduce the complexity of these terms, we consider the mean-field treatment, where the many-body interaction effects are mimicked via the interaction of a single electron system with a mean-field created by the aggregated effect of rest of the electrons. Mathematically speaking, such a two-body operator term $\hat{X}\hat{Y}$ can be approximated as

\begin{align}
 \hat{X}\hat{Y} &= \big\{\la\hat{X}\ra + (\hat{X} - \la\hat{X}\ra) \big\}\big\{\la\hat{Y}\ra + (\hat{Y} - \la\hat{Y}\ra)\big\}
 \notag\\
 &= \la \hat{X}\ra\la \hat{Y}\ra + \la \hat{X}\ra(\hat{Y}-\la\hat{Y}\ra)
 + \la \hat{Y}\ra(\hat{X}-\la\hat{X}\ra) + (\hat{X}-\la\hat{X}\ra)(\hat{Y}-\la\hat{Y}\ra)
 \notag\\
 &\approx \hat{X}\la\hat{Y}\ra + \hat{Y}\la\hat{X}\ra - \la\hat{X}\ra\la\hat{Y}\ra,
\end{align}

\noindent
where we have neglected the second order correction contribution. In practice, we can apply this idea of mean-field decoupling to the interaction terms of the Hamiltonian using different channels, such as density channel or pairing channel. Where density channel decoupling leads to the mean-fields of the form $\la c^{\dagger}c\ra$, and pairing channel decoupling leads to mean-fields of the form $\la c^{\dagger}c^{\dagger}\ra$. As we are specifically interested in superconducting solutions, we carefully choose pairing channel decoupling for our system and allow the superconducting correlation functions to take non-zero values to minimize energy.

Applying the mean field approximation described above on the interaction Hamiltonian (\ref{H_int_onsite_real}), and treating $\hat{X} = c_{i\upa}^{\dagger}c_{i\dna}^{\dagger}$ and $\hat{Y} = c_{i\dna}c_{i\upa}$, we get:

\begin{align}
 \mcal{H}_{\text{MF}}^{\text{onsite}} &=
 -U\summ{i} \left[ c_{i\upa}^{\dagger}c_{i\dna}^{\dagger}\la c_{i\dna}c_{i\upa}\ra 
 + c_{i\dna}c_{i\upa}\la c_{i\upa}^{\dagger}c_{i\dna}^{\dagger}\ra \right] +U\summ{i} \left[ \la c_{i\upa}^{\dagger}c_{i\dna}^{\dagger}\ra
 \la c_{i\dna}c_{i\upa}\ra \right] = -U\summ{i} 
 \left[ 
 \Delta_{i}c_{i\upa}^{\dagger}c_{i\dna}^{\dagger} + \text{H.c.} - |\Delta_{i}|^2 
 \right]
 \label{H_MF_onsite_real}
\end{align}
\noindent where, $\Delta_{i} = \la c_{i\dna}c_{i\upa}\ra$ is defined as the on-site pairing correlation at $i^{\text{th}}$ site.
Similarly, applying the mean field approximation to the terms $A$,$B$,$C$ and $D$ of Eq. (\ref{H_int_nn_real}), and decoupling into all possible pairing channels, we get:
\begin{align}
 \mcal{H}_{\text{MF}}^{\text{nn}} &=
 -V\summ{i,\delta} \left[ \Delta_{i\delta}^{\upa} c_{i\upa}^{\dagger}c_{i+\delta\upa}^{\dagger}
 + \text{H.c.} - |\Delta_{i\delta}^{\upa}|^2 \right] -V\summ{i,\delta} 
 \left[ 
 \Delta_{i\delta}^{\dna} c_{i\dna}^{\dagger}c_{i+\delta\dna}^{\dagger}
 + \text{H.c.} - |\Delta_{i\delta}^{\dna}|^2 
 \right] -V\summ{i,\delta} 
 \left[ 
 \Delta_{i\delta}^{+} c_{i\upa}^{\dagger}c_{i+\delta\dna}^{\dagger}
 + \text{H.c.} - |\Delta_{i\delta}^{+}|^2 
 \right]
 \notag\\
& \quad -V\summ{i,\delta} 
 \left[ 
 \Delta_{i\delta}^{-} c_{i+\delta\upa}^{\dagger}c_{i\dna}^{\dagger}
 + \text{H.c.} - |\Delta_{i\delta}^{-}|^2 
 \right]
 \notag\\
 \label{H_MF_nn_real}
\end{align}
\noindent where, we have defined the inter-site pairing correlations as:
\begin{align}
 \Delta_{i\delta}^{\upa} = \la c_{i+\delta\upa}c_{i\upa}\ra ;
 &\quad
 \Delta_{i\delta}^{\dna} = \la c_{i+\delta\dna}c_{i\dna}\ra
 \notag\\
 \Delta_{i\delta}^{+} = \la c_{i+\delta\dna}c_{i\upa}\ra ;
 &\quad
 \Delta_{i\delta}^{-} = \la c_{i\dna}c_{i+\delta\upa}\ra
 \label{pairing_correlations_real}
\end{align}
\noindent After combining all the terms together, our effective mean-field Hamiltonian in real space looks like:
\begin{align}
 \mcal{H}_{\text{MF}} &= -t\summ{\la ij\ra,\sigma}\left[ c_{i\sigma}^{\dagger}c_{j\sigma} + \text{H.c.}\right]
 -\mu\summ{i\sigma}c_{i\sigma}^{\dagger}c_{i\sigma} -B\summ{i}(c_{i\upa}^{\dagger}c_{i\upa}-c_{i\dna}^{\dagger}c_{i\dna}) -U\summ{i} 
 \left[ 
 \Delta_{i}c_{i\upa}^{\dagger}c_{i\dna}^{\dagger} + \text{H.c.} - |\Delta_{i}|^2 
 \right]
 \notag\\
 &\quad -V\summ{i,\delta} \left[ \Delta_{i\delta}^{\upa} c_{i\upa}^{\dagger}c_{i+\delta\upa}
 ^{\dagger}
 + \text{H.c.} - |\Delta_{i\delta}^{\upa}|^2 \right] -V\summ{i,\delta} 
 \left[ 
 \Delta_{i\delta}^{\dna} c_{i\dna}^{\dagger}c_{i+\delta\dna}^{\dagger}
 + \text{H.c.} - |\Delta_{i\delta}^{\dna}|^2 
 \right]
 \notag\\
& \quad -V\summ{i,\delta} 
 \left[ 
 \Delta_{i\delta}^{+} c_{i\upa}^{\dagger}c_{i+\delta\dna}^{\dagger}
 + \text{H.c.} - |\Delta_{i\delta}^{+}|^2 
 \right] -V\summ{i,\delta} 
 \left[ 
 \Delta_{i\delta}^{-} c_{i+\delta\upa}^{\dagger}c_{i\dna}^{\dagger}
 + \text{H.c.} - |\Delta_{i\delta}^{-}|^2 
 \right]
 \label{H_MF_total_real}
\end{align}
This Hamiltonian allows for spatially varying superconducting solutions since all the correlations are site dependent. However, working with the periodic boundaries and a uniform structure of superconducting correlation functions so that the translational invariance in preserved has a nice advantage. It allows us to block diagonalize the Hamiltonian by working in the Fourier space which drastically simplifies the problem. This is what we do in the next section.

\subsection{Effective Hamiltonian in Momentum Space}
\label{subsec_MF_momentum_space}
Observing the effective mean-field Hamiltonian in real space 
Eq. (\ref{H_MF_total_real}) it is apparent that the effective Hamiltonian
is a function of 
\begin{enumerate}
 \item a set of external parameters: t, $\mu$, $U$, $V$, $B$, and
 \item a set of complex superconducting configurations (pairing correlations):
 $\Delta_{i}$, $\Delta_{i\delta}^{+}$,
 $\Delta_{i\delta}^{-}$, $\Delta_{i\delta}^{\upa}$, and 
 $\Delta_{i\delta}^{\dna}$.
\end{enumerate}

In this article, we treat a clean system on a square lattice. Thus we exclude the possibility of inhomogeneity in our solutions, i.e., we expect our superconducting solutions to respect translational symmetry of the underlying lattice. This particular choice of translational symmetry makes Bloch basis to be more suitable to describe the effective Hamiltonian. Thus we apply the following transformation from Wannier basis to Bloch basis:
\begin{align}
c_{i\sigma} &= \frac{1}{\sqrt{N_{s}}}\summ{\vk} e^{-i\vk\cdot \vecr_{i}}
            c_{\vk\sigma}\enskip
            \text{ and }\enskip \notag\\
c_{i\sigma}^{\dagger} &= \frac{1}{\sqrt{N_{s}}}\summ{\vk} e^{i\vk\cdot \vecr_{i}}
            c_{\vk\sigma}^{\dagger}
\label{Ch2_fourier_transformation}
\end{align}
\noindent where $N_s$ is the total number of lattice points. This essentially block diagonalize the Hamiltonian and simplifies the problem. In Bloch basis, the effective mean-field Hamiltonian looks like:
\begin{align}
 \mcal{H}_{\text{MF}} 
 &= \summ{\vk}\left[\epsilon^{\upa}(\vk)c_{\vk\upa}^{\dagger}c_{\vk\upa} 
    + \epsilon^{\dna}(\vk)c_{\vk\dna}^{\dagger}c_{\vk\dna}\right] +\summ{\vk}\left[\Delta^{\upa\dna}(\vk)
    c_{\vk\upa}^{\dagger}c_{-\vk\dna}^{\dagger} + \text{H.c.} \right] +\summ{\vk}\left[\Delta^{\upa\upa}(\vk)
    c_{\vk\upa}^{\dagger}c_{-\vk\upa}^{\dagger} + \text{H.c.} \right]
 \notag\\
 & +\summ{\vk}\left[\Delta^{\dna\dna}(\vk)
    c_{\vk\dna}^{\dagger}c_{-\vk\dna}^{\dagger} + \text{H.c.} \right] + N_{s}U|\Delta|^2 + N_{s}V\summ{\delta}\left[ |\Delta_{\delta}^{+}|^2 + |\Delta_{\delta}^{-}|^2 
 + |\Delta_{\delta}^{\upa}|^2 + |\Delta_{\delta}^{\dna}|^2 \right]
\label{H_MF_total_kspace}
\end{align}
\noindent where, 
\begin{align}
 \epsilon^{\sigma}(\vk) &= -2t[\cos(k_x)+\cos(k_y)] -\mu- (-1)^\sigma B
 \notag\\
 \Delta^{\upa\dna}(\vk) &= -U\Delta -V\summ{\delta} 
 \left[ \Delta_{\delta}^{+}e^{-i(\vk\cdot\bdelta)} 
 + \Delta_{\delta}^{-}e^{i(\vk\cdot\bdelta)} \right]
 \notag\\
 \Delta^{\sigma\sigma}(\vk) &= 
 -V\summ{\delta}\Delta_{\delta}^{\sigma}e^{-i(\vk\cdot\bdelta)}
\label{H_MF_total_kspace_elements_definition}
 \end{align}

\noindent
here $\sigma={0,1}$ represent the spins $\{\upa,\dna\}$. To make use of translational invariance, we assumed spatial homogeneity of superconducting pairing correlations. So the set of variables: \{$\Delta_{i},\Delta_{i\delta}^{\upa},\Delta_{i\delta}^{\dna},\Delta_{i\delta}^{+}$ and $\Delta_{i\delta}^{-}$\} becomes independent of site index $i$ and our new set of variables becomes: \{$\Delta,\Delta_{\delta}^{\upa},\Delta_{\delta}^{\dna},\Delta_{\delta}^{+}$ and $\Delta_{\delta}^{-}$\}. In this effective Hamiltonian, $\epsilon^{\upa}(\vk)$ describes the kinetic energy of $\upa$-spin electrons with respect to an effective chemical potential $\mu_{\text{eff}}^{\upa} = \mu+B$. Similarly $\epsilon^{\dna}(\vk)$ describes the kinetic energy of $\dna$-spin electrons with respect to an effective chemical potential $\mu_{\text{eff}}^{\dna} = \mu-B$. It is evident that the presence of a magnetic field breaks the spin-degeneracy in the system via Zeeman coupling. It's worth mentioning that we work in the regime where magnetic field couples only with spin-degree of freedom. Orbital degree of freedom remains completely unperturbed by the presence of magnetic field, which leaves us with the freedom to ignore the Peierls substitution. In the above mentioned Hamiltonian Eq. (\ref{H_MF_total_kspace}) $\Delta^{\upa\upa}(\vk)$ and $\Delta^{\dna\dna}(\vk)$ represent superconducting correlations corresponding to equal spin pairing (ESP) states, while $\Delta^{\upa\dna}(\vk)$ represents superconducting correlations corresponding to opposite spin pairing (OSP) states. While the on-site attraction strength $U$ controls OSP states only, which is evident from the definitions of these pairing correlations Eq. (\ref{H_MF_total_kspace_elements_definition}), inter-site attraction $V$ controls both OSP and ESP states. This very fact suggests that $V$ is meant to play an important role in stabilizing non-trivial superconducting orders with variety of pairing symmetries.

\subsection{Bogoliubov-de Gennes Method}\label{sec:BdG}

The mean field Hamiltonian $\mcal{H}_{\text{MF}}$ Eq. (\ref{H_MF_total_kspace}) can be written
in terms of Nambu spinors ${\Psi}_{\vk}$ as:

\begin{align}
 \mcal{H}_{\text{MF}} = 
 \summ{\vk}{\Psi}_{\vk}^{\dagger}\mcal{H}_{\text{BdG}}(\vk){\Psi}_{\vk}
 \label{H_MF_nambu}
\end{align}
\noindent where, Nambu spinor ${\Psi}_{\vk}$ is a column vector of the form,
$ {\Psi}_{\vk} = (c^{}_{\vk\upa}\quad c_{-\vk\dna}^{\dagger}\quad 
c^{}_{\vk\dna}\quad c_{-\vk\upa}^{\dagger} )$
and $\mcal{H}_{\text{BdG}}(\vk)$ is a $4\times 4$ Hamiltonian matrix of the form:
\begin{align}
 \ham_{\text{BdG}}(\vk)=
 \begin{pmatrix}
          \eps^{\upa}(\vk) & \Delta^{\upa\dna}(\vk) &
          0 & \Delta_{s}^{\upa\upa}(\vk) \\
          \quad \\
          (\Delta^{\upa\dna}(\vk))^{\ast} & -\eps^{\dna}(-\vk) &
          (\Delta_{s}^{\dna\dna}(\vk))^{\ast} & 0 \\
          \quad \\
          0 & \Delta_{s}^{\dna\dna}(\vk) &
          \eps^{\dna}(\vk) & -(\Delta^{\upa\dna}(-\vk)) \\
          \quad \\
          (\Delta_{s}^{\upa\upa}(\vk))^{\ast} & 0 
          & -(\Delta^{\upa\dna}(-\vk))^{\ast} & -\eps^{\upa}(-\vk)
          \end{pmatrix}
\end{align}
$\mcal{H}_{\text{BdG}}(\vk)$ is known as the mean field Bogoliubov-de Gennes Hamiltonian.
We can diagonalize the BdG Hamiltonian by defining new Fermionic quasi-particle operators that mix the electronic operators as,
\begin{align}
 c_{\vk\upa} &= \sum^{'}_{\vk,\alpha} u_{\vk\upa}^{\alpha} \gamma_{\vk \alpha} 
      + (v_{\vk\upa}^{\alpha})^{\ast} \gamma_{-\vk \alpha}^{\dagger} 
      \notag \\
 c_{\vk\dna} &= \sum^{'}_{\vk,\alpha} u_{\vk\dna}^{\alpha} \gamma_{\vk \alpha} 
      + (v_{\vk\dna}^{\alpha})^{\ast} \gamma_{-\vk \alpha}^{\dagger}
\end{align}
\noindent The prime on the summation is a restriction to include only those states in the summation that will lead to a positive energy excitation spectrum of the diagonal Hamiltonian. Using this transformation, the Hamiltonian is diagonalized in the following form,
\begin{align}
 \mcal{H}_{\text{MF}} =  E_{g} + \summ{\vk \alpha}E_{\vk \alpha} 
 \gamma_{\vk \alpha}^{\dagger}\gamma_{\vk \alpha}
\end{align}
\noindent here $E_n$ is the excitation spectrum for the Bogoliubov quasiparticles. The physical constraint of non-negativity on the excitation energies is implemented by discarding the negative energy states from the definition of the Bogoliubov transformation. This results in $\{E_{\vk\alpha}\}$ to be a set of positive eigenvalues, and $\gamma_{\vk \alpha}^{\dagger}(\gamma_{\vk \alpha})$ creates (annihilates) a Bogoliubov quasiparticle with momentum $\vk$ and pseudo-spin $\alpha$. The BdG quasiparticles describe elementary excitations of the condensate, $E_{g}$ being the ground state energy of the condensate.

\subsection{d-vector formalism}\label{sec:d-vec}

In the absence of magnetic field, all the triplet superconducting solutions should be degenerate. However, while performing self-consistency, a randomized initial configuration should converge to a configuration with minimum energy. Due to degeneracy in the absence of magnetic field, whichever initial triplet configuration we provide, the self-consistency approach tends to throw out an arbitrary configuration. To bypass this issue, we divide all the superconducting configurations into different classes based on the characteristics we are most interested in, namely, singlet and triplet pairing. We do so by employing the $d$-vector formalism.

The most general superconducting gap function can be represented in a matrix form as,
\begin{eqnarray}
 \Delta({\bf k}) = \begin{pmatrix}
 \Delta^{\uparrow\uparrow}({\bf k})& \Delta^{\uparrow\downarrow}({\bf k})\\  \Delta^{\downarrow\uparrow}({\bf k}) & \Delta^{\downarrow\downarrow}({\bf k})
 \end{pmatrix}
\end{eqnarray}
In this, the elements correspond to the spin state of the electrons that constitute the Cooper pair. Furthermore, we may write the Cooper pair wavefunction as,
\begin{eqnarray}
 \psi_{\sigma\sigma'}({\bf k}) = [ \Delta_0({\bf k}) + {\bf d}({\bf k}) \cdot {\bf \sigma} ]i(\sigma_y)_{\sigma\sigma'}
\end{eqnarray}
Where, $\psi_{\sigma\sigma'}(k)$ is the Cooper pair wavefunction with $\sigma \sigma'$ pairing. The first term, $\Delta_0({\bf k})$, is the superconducting gap function for the singlet type of superconductor, whereas, the three-components of the triplet superconducting gap function is related to the complex vector ${\bf d}({\bf k})$. The advantage of writing it in this form is that a rotation of spin quantization axis in spin space would be equivalent to a 3D rotation of ${\bf d}({\bf k})$ vector. A rotation of ${\bf d}({\bf k})$ vector, would in turn adjust the superconducting gap functions corresponding to the three spin triplet components accordingly, and therefore, it makes it easier to track them. Now, since the length of ${\bf d}({\bf k})$ vector (and averaged over ${\bf k}$ space) will remain invariant of spin-quantization axis, we may use this quantity to track down the overall triplet component of superconductivity. We therefore use,
\begin{eqnarray}
 \Delta_{\text{SP}} &= \frac{1}{N_s} \Sigma_{\bf k} |\Delta_0({\bf k})| \\
 \Delta_{\text{TP}} & = \frac{1}{N_s} \Sigma_{\bf k} |{\bf d}({\bf k})|
\end{eqnarray}
as our definitions for the singlet superconducting gap function and triplet superconducting gap functions, respectively.

This formalism also gives us an handle on several properties of the superconducting state for example unitarity of the superconducting wavefunction and time reversal symmetry. A superconducting gap function, in matrix form, $\Delta({\bf k})$ is called unitary if the product $\Delta({\bf k})\Delta^{\dagger}({\bf k})$ is proportional to the unit matrix, otherwise the superconducting gap function, in matrix form, is known as non-unitary. With this definition, it is clear that only triplet pairing matrices can be non-unitary since,
\begin{eqnarray}
\Delta^{\dagger}({\bf k})\Delta({\bf k}) = |{\bf d({\bf k})}|^2 \mathbb{I} + \bf{q} \cdot \bf{\sigma} 
\end{eqnarray}  
where ${\bf{q}} = i (\bf{d}\times \bf{d}^*)$. The measure of $\Delta_{\bf q}\equiv \sum_{\bf k}|{\bf q}| $ tells us if the superconducting phase is unitary. Also, the time reversal symmetry can easily be checked by,
\begin{eqnarray}
{\bf d}({\bf k}) \stackrel{?}{=}  -{\bf d}({\bf k})^* ~~~~~~~~~~~~~~\text{(Time reversal invariant if equal)}
\end{eqnarray}  
Therefore, using the $d$-vector approach we can infer many properties of the superconducting wavefunction. 

\subsection{self-consistency and minimization}

In this section we lay out how the method of self-consistency is implemented computationally. We set electronic hopping parameter $t = 1$ as the basic energy scale, then we are left with four independent external parameters in the Hamiltonian, {\it viz.}, $\mu, U, V, B$. Corresponding to these, we will obtain a set of self-consistent SC pairing correlations, $\{\Delta\}$ that defines the SC OP. To solve the BdG equations numerically, an initial guess of $\{\Delta\}$ is fed into the Hamiltonian at some external parameter value. This Hamiltonian is diagonalized and eigenvalues and eigenvectors are calculated. The obtained eigenspectrum is used to further redefine the Hamiltonian and is re-diagonalized. This set of steps is labelled as an iteration and the cycle of iteration is repeated until the averages converge within a specified error, which was set to $10^{-5}$ in our calculations. The ground state (at $T=0$) energy is calculated by summing over all the positive energy states. Therefore, the problem now reduces to minimizing the total energy w.r.t. the set $\{\Delta\}$ of pairing correlations. Since we do not fix any constraint on the nature of the pairing correlation, energy minimization shall result in the most energetically favourable pairing correlation symmetry that may either by singlet like or triplet like or even some strange combination of these two. This method of calculating averages self-consistently is actually equivalent to the energy minimization.

\bibliography{IntroBib}


\section*{Acknowledgements}

We acknowledge the use of Computing Facility at IISER Mohali.

\section*{Author contributions statement}

S.K. and S.N. conceived the project. S.N. and N.B. formulated the methodology and ran preliminary simulations for consistency. 
S.N. performed the simulations and arranged the results, as presented in the final manuscript. All authors contributed to analyzing the results and writing the final manuscript. S.K. supervised the project.

\section*{Additional information}
All the codes are written from scratch, in Python or Fortran. Plots are generated using Python scripts.

\end{document}